# WeSME: Uncovering Mutual Exclusivity of Cancer Drivers and Beyond

Yoo-Ah Kim[1], Sanna Madan[2], and Teresa M. Przytycka[1] *

[1] National Center for Biotechnology Information, NLM, NIH, Bethesda, MD 20894

[2] Poolesville High School, Poolesville, MD

*To whom correspondence should be addressed; przytyck@ncbi.nlm.nih.gov

**Abstract**

Mutual exclusivity is a widely recognized property of many cancer drivers. Knowledge about these relationships can provide important insights into cancer drivers, cancer-driving pathways, and cancer subtypes. It can also be used to predict new functional interactions between cancer driving genes and uncover novel cancer drivers. Currently, most of mutual exclusivity analyses are preformed focusing on a limited set of genes in part due to the computational cost required to rigorously compute p-values. To reduce the computing cost and perform less restricted mutual exclusivity analysis, we developed an efficient method to estimate p-values while controlling the mutation rates of individual patients and genes similar to the permutation test. A comprehensive mutual exclusivity analysis allowed us to uncover mutually exclusive pairs, some of which may have relatively low mutation rates. These pairs often included likely cancer drivers that have been missed in previous analyses. More importantly, our results demonstrated that mutual exclusivity can also provide information that goes beyond the interactions between cancer drivers and can, for example, elucidate different mutagenic processes in different cancer groups. In particular, including frequently mutated, long genes such as TTN in our analysis allowed us to observe interesting patterns of APOBEC activity in breast cancer and identify a set of related driver genes that are highly predictive of patient survival. In addition, we utilized our mutual exclusivity analysis in support of a previously proposed model where APOBEC activity is the underlying process that causes TP53 mutations in a subset of breast cancer cases.

*Accessibility*: http://www.ncbi.nlm.nih.gov/CBBresearch/Przytycka/index.cgi#wesme

## 1 INTRODUCTION

Mutual exclusivity is a frequently observed property in two or more genes having mutations in cancer patients [1-8]. Mutually exclusive gene pairs (i.e., gene pairs of which simultaneous mutations in the same patients are less frequent than it is expected by chance) are observed within many cancer types, while some are common across multiple different cancer types [1]. The underlying cause of mutual exclusivity might vary depending on the context. In many cases, genes mutated in a mutually exclusive way have been found to be members of the same functional pathway. In such a case, the mutual exclusivity of their mutation pattern can be explained by the pathway centric view of cancer. Specifically, if two genes belong to the same cancer-driving pathway, then a mutation in just one of them might suffice to dysregulate the pathway and trigger cancer progression [2, 5, 7, 9-11]. In contrast, in a combined analysis of multiple cancer types, mutually exclusive aberrations might represent tissue specific cancer drivers. In fact, the sets of mutually exclusive drivers obtained in Pan-Cancer analysis [12, 13] predominantly included the tissue type specific genes that, typically, do not share common pathways. Yet, even within the same tissue type, mutual exclusivity between genes from different pathways is not unusual [1]. This type of exclusivity can be explained, for example, by synthetic lethality where simultaneous mutations in both genes are detrimental for cancer progression [14, 15]. Another (not necessarily unrelated) possibility is that the mutually exclusive drivers define different cancer subtypes within the same tissue type. Finally, mutual exclusivity can also be observed between less frequently mutated cancer drivers suggesting more subtle relationships between them and their contribution to cancer progression. For example, previous research found that even if the exclusivity between ARID1B and KRAS is not significant in any of individual cancer types separately, the pair was shown to be mutually exclusive when analyzing all cancer types together as statistical power was gained [1].

Mutual exclusivity can potentially provide important information about cancer, including its alternative progression pathways and subtypes. It can also reveal important relationships between genes and pathways. However, most of the analyses of mutual exclusivity so far have been focused on limited subsets of genes that were selected based on mutation rate cut-off (typically 3% or 5% depending on datasets) or restricted by other criteria. One of the factors limiting the applications of mutual exclusivity analysis beyond restricted gene sets was the cost of computing rigorous p-values for all gene pairs of interest. Statistically, pairwise mutual exclusivity can be estimated by a permutation test in which the original mutational profile is permuted by swapping random pairs of mutations iteratively while preserving the mutation rates of genes and patients. A possibly faster alternative approach is to use a hypergeometric test. However, the hypergeometric test treats all patients equally, ignoring patient specific mutation frequencies, which may limit its accuracy.

To provide a more in-depth understanding of mutual exclusivity in cancer data, we started by designing a new and fast heuristic for estimating statistical significance of mutual exclusivity relationships. Our method, WeSME (We̲ighted S̲ampling based M̲utual E̲xclusivity), estimates p-values of mutual exclusivity while taking into account mutation frequencies of patients. It closely approximates the results of the permutation-based method and does so without using the costly permutations of a mutation matrix. Moreover, WeSME can compute p-values for a subset of genes independently from the rest of genome unlike the permutation method that requires whole genome permutations. By dynamically adjusting sampling depth, WeSME can provide high precision p-values without a significant increase in computational cost. Adopting the same sampling technique, we also developed a complementary test, WeSCO (Weighted Sampling based Co-Occurrence), to estimate the significance of co-occurrence in gene mutations.

With WeSME at hand, we began to explore the information that a comprehensive mutual exclusivity survey can provide. We included all genes with mutations in the analysis without pre-filtering. In particular, we did not exclude very long genes such as muscle protein Titin (TTN), gel forming Mucins, which often emerge as frequently mutated genes in cancer (even after correction with their lengths) but are generally assumed to be artifacts [16]. Interestingly, we found that some of these genes have many mutual exclusivity partners. In particular, we identified TTN as one of the hubs in the mutual exclusivity network in breast cancer and endometrial carcinoma. The high number of mutual exclusivity partners of TTN prompted us to hypothesize that mutual exclusivity of its mutations could be a reflection of an underlying mutagenic process that occurs in a specific subgroup of patients.

To test this hypothesis, we reasoned that long and frequently mutated genes might harbor a recognizable signature of the mutagenic process acting on them. Specifically, it has been recently recognized that a subclass of APOBEC cytidine deaminases, which convert cytosine to uracil during RNA editing, are a source of mutagenicity in human tumors [16-18]. Several other factors such as Pol ε mutation, patient ages at diagnosis, known mutagenic exposures, or defects in DNA maintenance are also associated with characteristic mutation signatures [19]. We found that, in the case of breast cancer, the mutation pattern of TTN is consistent with the APOBEC3B activity, while in endometrial carcinoma, it is consistent with the Pol ε mutation signature.

Including TTN gene in our mutual exclusivity analysis allowed us not only to observe interesting patterns of APOBEC activity in breast cancer but also to identify a set of driver genes that is highly predictive of patient survival. In addition, we showed that our mutual exclusivity analysis supports the model where APOBEC activity is the underlying process that causes TP53 mutations in highly mutated breast cancer cases. Finally, with our method, we have been able to uncover mutually exclusive cancer drivers and novel putative drivers despite their relatively low mutation rates in the breast cancer dataset.

Taken together, our results demonstrated the utility of our fast statistical test for mutual exclusivity and showed that the unrestricted analysis of mutual exclusivity enabled by the method can provide valuable biological insights.

## 2 WeSME (WEIGHTED SAMPLING BASED MUTUAL EXCLUSIVITY) METHOD

Both hypergeometric and permutation based methods have been popular choices for estimating the significance of mutual exclusivity. The drawback of the permutation-based method is the computational cost, thus limiting gene sets to be considered and/or precision. On the other hand, the hypergeometric test does not take into account different mutation frequencies of patients and often underestimates the significance of exclusivity. WeSME aims to reduce the computational cost of estimating the significance of mutual exclusivity by utilizing weighted samplings instead of performing costly permutations of a mutation matrix, while accurately computing p-values by including patient mutation frequency information in sampling probability (See the discussion on the computational costs in Supplementary Section S-A).

### 2.1 General description of WeSME method

WeSME first computes the mutation frequencies of patients based on the observed distribution (Algorithm 1 and Figure 1A). Let $G = \{g_1, g_2...g_m\}$ and $S = \{s_1, s_2 ... s_n\}$ be the set of all genes and cancer patients, respectively. Let $mf(s)$ be the observed mutation frequency of a patient $s$. Next we obtain the null distribution of the mutation pattern of a gene by performing samplings of patients based on the weights $(mf(s_1), mf(s_2) ... mf(s_n))$ (Algorithm 2). Specifically, for a gene $g$ in $G$ with $m(g)$ mutated patients, we randomly sample $m(g)$ patients without replacement while making sure that the probability of a patient $s$ being sampled is proportional to the observed mutation frequency $mf(s)$ (i.e., sampling weighted by $mf(s)$). Let $(R_g^1, R_g^2, R_g^3 ... R_g^W)$ denote the sampling results when $W$ weighted samplings were performed for gene $g$. $R_g^l$ corresponds to the random subset of patients for gene $g$ in the $l$-th weighted sampling and the size of $R_g^l$ must be $m(g)$ for all $l$. Note that the samplings depend only on $m(g)$ and W, and therefore two genes $g_1$ and $g_2$ with $m(g_1)=m(g_2)$ can reuse the same samplings. We use this fact to further reduce the sampling time (see Section 2.2 and Supplementary Section S-A).

To estimate the p-value of mutual exclusivity of two genes $g_i$ and $g_j$ in $G$, we compare the number of patients exclusively covered by the two genes and the numbers of such patients from the null distribution (Algorithm 3). Formally, for two subsets of patients $S_i$ and $S_j$, let us define $ExCover(S_i, S_j)$ to be $|XOR(S_i, S_j)|$. Suppose that $S(g_i)$ is the set of patients who have mutations in gene $g_i$. Then $ExCover(S(g_i), S(g_j))$ represents the exclusive cover size by either gene $g_i$ or $g_j$. As in the permutation-based method, the empirical *p*-value of mutual exclusivity of $g_i$ and $g_j$ is given as

$$p(g_i, g_j) = |\{l=1\ldots W \mid ExCover(R_i^l, R_j^l) > ExCover(S(g_i), S(g_j))\}|/W$$

where $(R_i^1, R_i^2, R_i^3 \ldots R_i^W)$ and $(R_j^1, R_j^2, R_j^3 \ldots R_j^W)$ are the random samplings of $g_i$ and $g_j$, respectively. We note that the precision of *p*-values depends on the number of weighted sampling pairs *W*. In Subsection 2.2, we discuss how we increase W to obtain a better precision without increasing the computational time significantly.

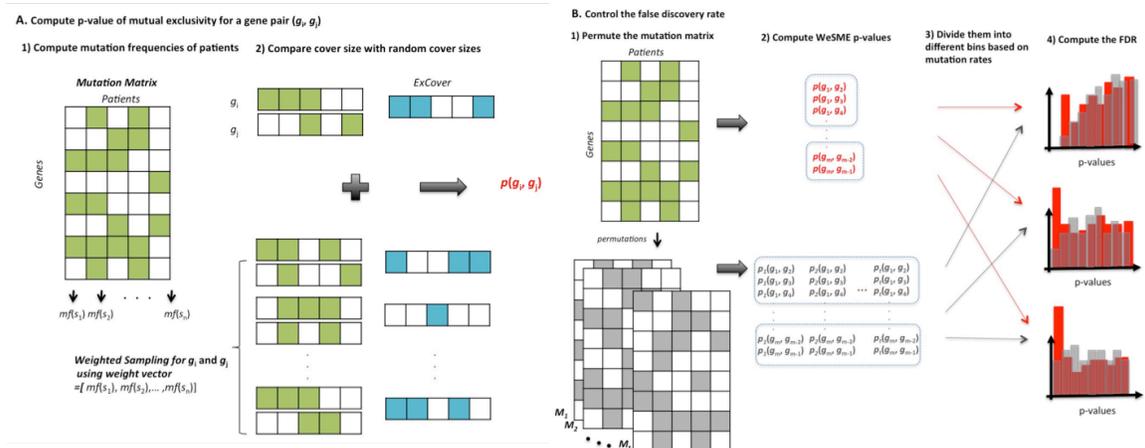

**Figure 1.** Illustration of WeSME method

One noteworthy advantage of using WeSME over the permutation method is that except Algorithm 1, the analysis can be carried out for a subset of genes independently from the rest of genome whereas in the permutation method, the whole genome permutations need to be performed to obtain random profiles.

### 2.2 WeSME provides a fast and accurate approximation to the permutation based method

In this subsection, we discuss some optimization techniques to further reduce the computational costs of WeSME and present the results applied to TCGA BRCA dataset.

**Compact sampling**: given the observed mutation frequency vector $(mf(s_1), mf(s_2) \ldots mf(s_n))$ of patients, the weighted sampling of a gene *g* depends only on the number of mutated patients *m(g)* and the number of samplings W. Utilizing this fact, we ran Algorithm 2 for each different *m(g)* instead of running for each gene. More specifically, let *K* be $\{k \mid m(g) = k \text{ for some } g\}$. We performed W weighted samplings for each different *k* in *K* (denoted as *WS(k)*). As discussed in Supplementary Section S-A, we found that it reduces the sampling time and storage requirement significantly.

**Dynamic depth of precision**: Suppose that we run Algorithm 3 for two genes $g_i$, $g_j$. Let $k_i = m(g_i)$ and $k_j = m(g_j)$ and let $(R_i^1, R_i^2, R_i^3 \ldots R_i^W)$ and $(R_j^1, R_j^2, R_j^3 \ldots R_j^W)$ denote the weighted random samplings for $k_i$ and $k_j$ respectively. Then there are at least $W(W-1)$ different ways to create the pairs (discarding $R_i^l$ and $R_j^l$ when $k_i = k_j$ in which case $R_i^l$ and $R_j^l$ are identical) and compute *ExCover*. For added randomness and optimizing the running time, we started by randomly selecting a small number of pairs ($W_0$) and compute *ExCover* for them initially. For example, we start with $W_0 = 1,000$ pairs at first and if empirical p-value < 0.1 (there are less than 100 random pairs whose *ExCover* is bigger than the original *ExCover*), we increased the number of sampling pairs by 10 times to obtain a better precision. The maximum precision we used in this analysis is $10^{-6}$ (i.e., the maximum number of pairs $= 10^6 \leq W(W-1)$)

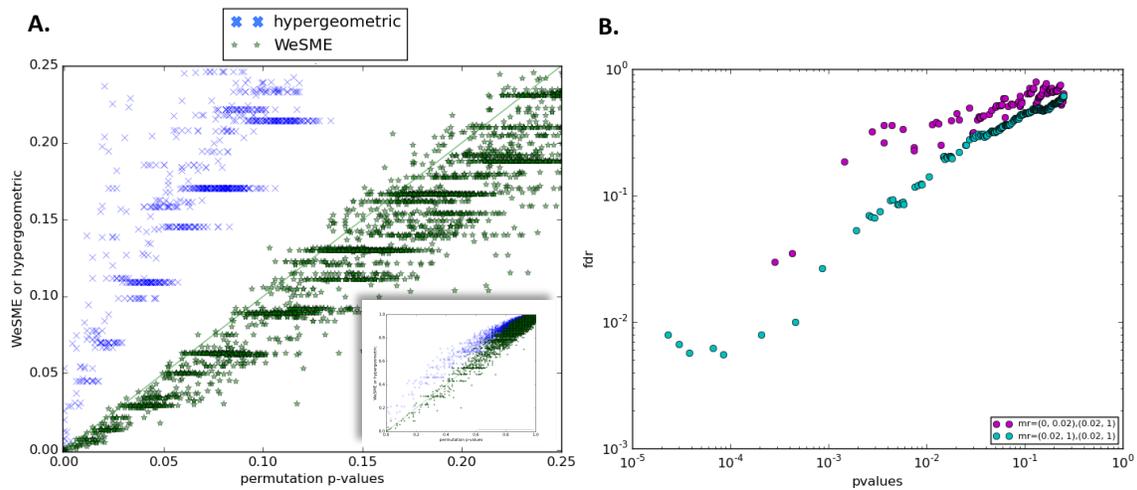

**Figure 2** (A) Comparison of the p-values computed based on permutation, hyergometric and WeSME tests. Only the data for the pairs with their p-values ≤ 0.25 are shown. The p-values for the gene pairs in HumanNet are shown in the inset. (B) The plot shows the WeSME p-values (p ≤ 0.25) and their corresponding FDRs. FDRs are computed separately depending on the mutation rates of genes.

**Analysis of TCGA BRCA somatic mutations**: we ran WeSME with the somatic mutation profile of 665 TCGA BRCA patients. We considered all genes with at least one mutation without pre-filtering and computed the p-values of all gene pairs. Figure 2A shows the comparison of the p-value distributions obtained by WesME, permutation test, and hypergeometric test for the gene pairs with their p-values ≤ 0.25. The p-values of the permutation-based method were computed using 1,000 randomly permuted mutation profiles (permuted while preserving the mutation frequencies of patients and the mutation rates of genes). We found that the results obtained with WeSME approximate the ones with the permutation test very well (relative error rate ≅ 0.17, Pearson corr. coef. ≅ 0.98) while the hypergeometric test, in general, underestimates the mutual exclusivity (relative error rate ≅ 0.97, Pearson corr. coef. ≅ 0.89). We noted that WeSME p-values tend to be slightly more significant than the p-values based on the permutation test especially for the gene pairs with significant p-values (see more discussion in Supplementary Section S-C). Still WeSME approximates the permutation test much better than the hypergeometric test for the significant gene pairs (for pairs with p-value ≤ 0.05, relative error rate – 0.30 vs. 3.3, Pearson corr. coef. – 0.89 vs. 0.76). This trend of over-estimation was less prominent when we considered all gene pairs in HumanNet[1] (Figure 2A inset, the whole spectrum of p-values are shown).

**False discovery rate (FDR)**: in order to control FDR, we obtained the null distribution by generating 100 permuted mutational profiles and computing WeSME p-values for all gene pairs with the permuted mutations (Figure 1B and Algorithm 4). Observing that the null distribution of p-values showed different patterns depending on the mutation rates of involved genes, we divided the genes into two groups based on their mutation rates (highly or rarely mutated), which in turn divided all gene pairs into three different bins. The null distribution for each bin was created separately using the p-values from the permuted mutation profiles. We then computed the FDR of the p-value for the gene pairs in each bin by comparing them with the null distribution for the corresponding bin. Figure 2B shows the different FDRs obtained using three bins – (R, R), (R, H), and (H, H) for BRCA where R (*R*arely mutated) includes all genes with the mutation rate of 2% or lower, and the remaining genes are in H (*H*ighly mutated). There were no pairs in bin (R, R) with p-value ≤ 0.25. In Figure 2B, we can observe that the pairs in bin (R, H) have higher FDRs (shown in purple) than the pairs in bin (H, H) (shown in cyan) for the same p-values.

---

[1] http://www.functionalnet.org/humannet/

## 2.3 Mutual exclusivity network in BRCA and UCEC

Figure 3A shows the network with significant mutually exclusive (ME) edges created with BRCA somatic mutation profiles. We included the gene pairs with p value ≤ 0.01 for highly mutated gene pairs in (H, H) and applied a stricter p-value threshold of 0.001 for pairs in (R, H). The corresponding FDR ≤ 0.125. We also computed the ME network with TCGA endometrial carcinoma (UCEC) somatic mutation data with a stricter FDR cut-off of 0.0025 (Figure 3B). The more stringent FDR cutoff was used because the analysis of UCEC patients resulted in a larger number of ME pairs, presumably due to its partition into subtypes with drastically different mutational profiles. Additional analysis with TCGA Acute Myeloid Leukemia dataset is included in Supplementary Section S-E.

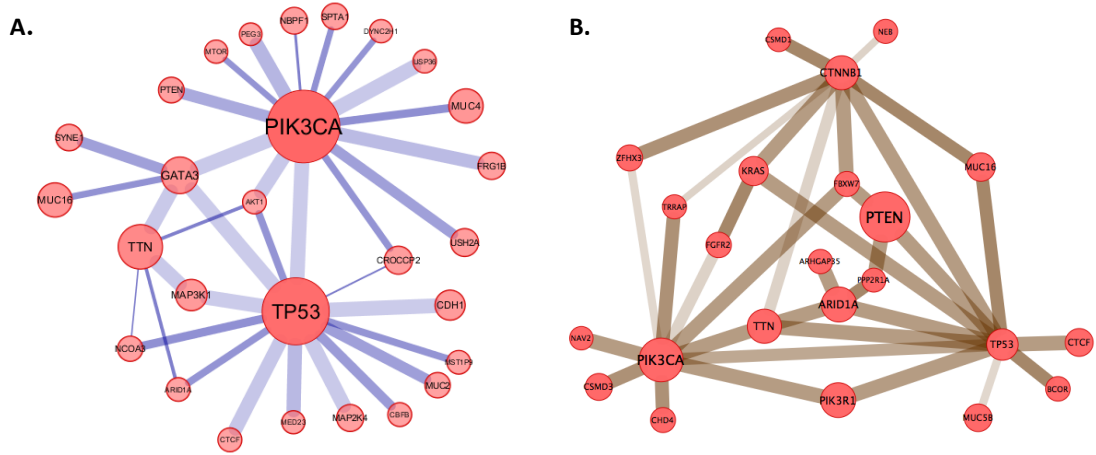

**Figure 3.** ME networks for (A) BRCA (FDR ≤ 0.125) and (B) UCEC (FDR ≤ 0.0025). The node size of each gene represents the mutation rate of the gene for the given cancer type. The widths of edges correspond to their WeSME p-values (thicker edges for more significant pairs) whereas darker colors of the edges correspond to lower FDRs. The networks are created using Cytoscape [20]

The BRCA network (Figure 3A) contains 34 mutually exclusive gene pairs. There were 13 pairs where one of the genes in the pair was mutated in 3% or less patients including two pairs where the mutation rate was below 2%. These genes would have been filtered out if 3% of mutation rate threshold was applied. Interestingly, these pairs include five pairs of well-known drivers: (ARID1A, TP53), (MTOR, PIK3CA), (AKT1, PIK3CA), (AKT1, TP53), (CTCF, TP53); four pairs where one member is a known cancer driver and its partner is a likely (and previously implicated) cancer driver: (TP53, MED23), (PIK3CA, PEG3)[21], (CBFB, TP53) [22] (USP36, PIK3CA) [23]; and three pairs where one member is a long gene - Titin (TTN) or Cytoplasmic Dynein 2 Heavy Chain (DYNC2H1) – while its partner is a known cancer driver (TTN, ARID1A), (AKT1, TTN), (DYNC2H1, PIK3CA). Previous work suggested that TTN and DYNC2H1 are unlikely to be cancer drivers [24]. Yet, TTN has rather a large number of partners in the mutual exclusivity network, suggesting that patients carrying mutations in this gene may have some distinguishing properties.

As for endometrial carcinoma, the previous analysis on TCGA UCEC dataset identified 12 genes (PTEN, TP53, PIK3CA, PIK3R1, ARID1A, ARID5B, KRAS, CTCF, CTNNB1, FBXW7, PPP2R1A, RPL22) as recurrently mutated in different subtypes [25]. All but ARID5B and RPL22 are included in the ME network obtained with FDR cut-off of 0.0025. RPL22 and TP53 pair had FDR=0.00256 while the most significant ME partner of ARID5B was CTNNB1, for which FDR=0.02. Though overlooked in the TCGA publication, TRRAP (transformation/transcription domain-associated protein) is likely to be an important driver. The gene encodes a protein that plays a role in transcription and DNA repair. With our very stringent FDR cut-off, TRRAP is mutually exclusive with CTNNB1 and PIK3CA. Finally, as in the case of BRCA, the TTN gene has four neighbors, all of which are prominent cancer drivers.

In our subsequent analysis in Section 3, we show that although TTN may not be a cancer driver itself, it is likely to be a passenger of a mutagenic process that is specific to a subgroup of patients in these two cancer types.

## 3 BEYOND MUTUAL EXLUSIVITY OF DRIVERS - DISCOVERING DRIVER SIGNATURES FROM MUTUAL EXLUSIVITY PARTNERS OF PASSENGERS

### 3.1 APOBEC and POL ε mutational signatures of TTN

Recent studies identified several genes that are highly mutated but still unlikely to be cancer drivers [16]. These genes are often extremely long genes, including the muscle protein Titin (TTN), gel forming mucins (MUC), and ryanodine receptors (RYR1). Those genes were mostly excluded from candidate cancer driver lists in subsequent analyses. However, some of the apparent false-positive findings may not be simply caused by their lengths since the methods for identifying significantly mutated genes typically account for gene length information.

TTN is one of the genes identified as the ones that are not likely a cancer driver despite its high mutation rate but the prominent position of TTN in both BRCA and UCEC mutual exclusivity network strongly suggests that the mutagenesis of TTN may provide valuable information about patient subgroups. That is, even if TTN is not a cancer driver, the gene may be a passenger that is mutated in a specific group of patients and be a "witness" of a mutagenic process occurring in the subgroup.

To confirm the hypothesis that TTN is a witness of a particular mutational process, we analyzed the mutational spectrum of this gene for both cancer types (BRCA and UCEC), and compared it to the signatures of mutational processes in human cancer [19]. We found that the mutational signature of TTN in BRCA dataset was consistent with the APOBEC signature. It showed a significant bias towards C>T and C>G mutations – 67 and 44 mutations respectively (38% and 25% of 175 mutations in total and 50% and 33% of 132 C cite mutations). In addition, most of these mutations are in the context where C is preceded by A or T (71%).

In UCEC dataset, on the other hand, TTN has a different mutation signature, which is consistent with Pol ε mutations. TTN showed a strong preference for C>A and C>T mutations, having 231 and 347 mutations respectively, which represents 30% and 46% of 761 mutations in total and 40% and 59% of C cite mutations. Moreover, 123 of C>A mutations (53%) are TCT>TAT and 116 mutations of C>T mutations (33%) are TCG>TTG, strongly demonstrating that TTN mutations in UCEC is associated with Pol ε mutation signature.

### 3.2 Mutual exclusivity patterns of TTN and TP53 indicate a group of genes predictive of patient survival and suggest a role of APOBEC3 activity in TP53 mutations.

**ME network and survival analysis**: An interesting property of mutational landscape of breast cancer is that frequently mutated genes such as TP53, PIK3CA, TTN and GATA3 in breast cancer do not have a strong discriminative power of patient survival (only TP53 mutations are weakly discriminative, p-value = 0.08). We therefore asked the question whether the mutual exclusivity network could suggest a set of genes with a better discriminative power.

Previous studies have suggested that APOBEC3B activity correlates with unfavorable prognosis in breast cancer. For example, it has been shown that APOBEC3B gene expression correlates with the Genomic Grade Index (GGI) score and that high APOBEC3B gene expression was associated with recurrence after treatment [26]. According to a separate study, APOBEC3B was associated with poor survival in ER+ breast cancer patients [27]. Building on our observation that TTN mutations are largely due to APOBEC3B activity (as discussed in the previous section), we hypothesized that putative cancer drivers exclusive with TTN mutations would be characterized by better prognosis. This was indeed confirmed in our survival analysis. We found that four genes that are exclusive with TTN (MAPK3K1, NCOA3, AKT1, ARID1A) were a good predictor of survival jointly and also predict disease free survival after treatment very well (Figure 4B). Interestingly, TTN and TP53 share the discriminative gene set as their direct ME neighbors in the BRCA ME network (Figure 3A). On the other hand, adding other putative drivers exclusive with TP53 but not with TTN (FOX1A, MED23, MAP2K4)[2] reduced the discriminatory power of the gene set.

**Role of APOBEC3B activity in TP53 mutations**: The similarity of exclusivity patterns of TTN and TP53 is consistent with the recently discovered relationship between APOBEC enzymatic activity and TP53 mutations [18]. In their research, Burns *et al*. hypothesized that APOBEC3B-catalysed deamination provides a chronic source of DNA damage that could cause TP53 mutations.

To support the model proposed by Burns et al., we need to test if it is possible that the mutations of both genes in BRCA depend on APOBEC activity but are independent of each other after controlling for their dependency on APOBEC. That is, under this model, we expect TTN and TP53 to co-occur but the significance of co-occurrence to disappear after correcting with mutation frequencies of patients. To test if this is the case, we developed WeSCO (Weighted Sampling based Co-Occurrence) by adopting the same sampling technique as WeSME for estimating the significance of co-occurrence of mutations corrected with patient mutation frequency. Indeed, we did not find a significant co-occurrence (Figure 4C, p-value = 0.29) of TTN and TP53 when applying WeSCO test while the co-occurrence was highly significant without the correction (p-value = 0.0002, hypergeometric test).

Although it is currently not known how APOBEC activity leads to TP53 mutations, we conjecture that immune response might be the precursor step leading, via APOBEC activity, to TP53 mutation for some cancer patients. It has been demonstrated that the AID/APOBEC family plays important roles in adaptive and innate immunity [28] and APOBEC up-regulation is correlated with HPV-positive status, which implicates an HPV-mediated mechanism of APOBEC3B up-regulation for these cancer types [29]. In particular, the authors provided evidence that APOBEC activity is responsible for the generation of helical domain hot spot mutations in PIK3CA across multiple cancers. The key role of APOBEC family in natural and adaptive immunity strongly suggests that APOBEC up-regulation may be also related to immune response in non-HPV mediated cases and our results support a similar mechanism for TP53 mutations in some breast cancer cases (but not all cases as discussed below).

**Underlying causes of TP53 mutations may be heterogeneous**: We observed that the subgroup of patients with TP53 mutations but no TTN mutations had significantly lower mutation

---

[2] CTCF was not included because it shows a slight tendency of exclusivity with TTN, p-value < 0.15

frequencies compared to the subgroup of patients with TTN mutations (Figure 4C; p-value = 0.025). This suggests that the TP53 only group also contains patients whose TP53 mutations are not due to APOBEC activity as otherwise we would not expect to see a difference in mutation frequencies between the TP53 only and the TTN patient groups. The difference of mutation frequencies in the group suggests that underlying causes of TP53 mutations in breast cancer are heterogeneous.

Interestingly, TP53 and TTN mutations are mutually exclusive in the analysis of endometrial carcinoma data where TTN mutations show Pol ε mutation signature (Figure 3B). This observation suggests that the mutations in TP53 are generally not caused by the mutagenic process related to Pol ε mutations.

In summary, our results are consistent with the previously proposed model in which APOBEC3B can be a source of mutations in TP53 for highly mutated cancer cases, however reject the model in which this is true across all breast cancer patients with TP53 mutations.

## 4 CONCLUSIONS

We performed a whole genome scale analysis of mutual exclusivity of genes with mutations in breast cancer and endometrial carcinoma. For this purpose, we first developed WeSME - a new, fast method to estimate statistical significance of mutual exclusivity. Similar to the permutation based method, WeSME controls mutation frequencies of patients and mutation rates of genes in the null model, but can be performed quickly with a nearly arbitrary level of precision. The p-values estimated by WeSME closely matched the p-values obtained with the permutation-based method. In addition, our proposed method to control FDR takes into account the differences of p-value distributions from gene pairs with different mutation rates.

The development of WeSME allowed us to perform a large-scale mutual exclusivity analysis without restricting it to highly mutated genes or genes known *a priori* to be cancer drivers. The approach led us to uncover mutually exclusive cancer drivers, some of which have relatively low mutation rates. In addition, our mutual exclusivity analysis allowed us to pinpoint differences between mutagenic processes in different cancer groups. In particular, the inclusion of TTN (a very long and frequently mutated gene) not only allowed us to observe the patterns of APOBEC activity in breast cancer and Pol ε mutation signature in endometrial carcinoma but also help identify a set of driver genes that is highly predictive of patient survival. TTN was considered to be an artifact and typically filtered out in cancer analyses.

Taken together, we developed a tool enabling a fast genome-wide analysis of mutual exclusivity of mutations in cancer and used it to demonstrate that the genome-wide mutual exclusivity analysis can provide information beyond relationships between cancer drivers.

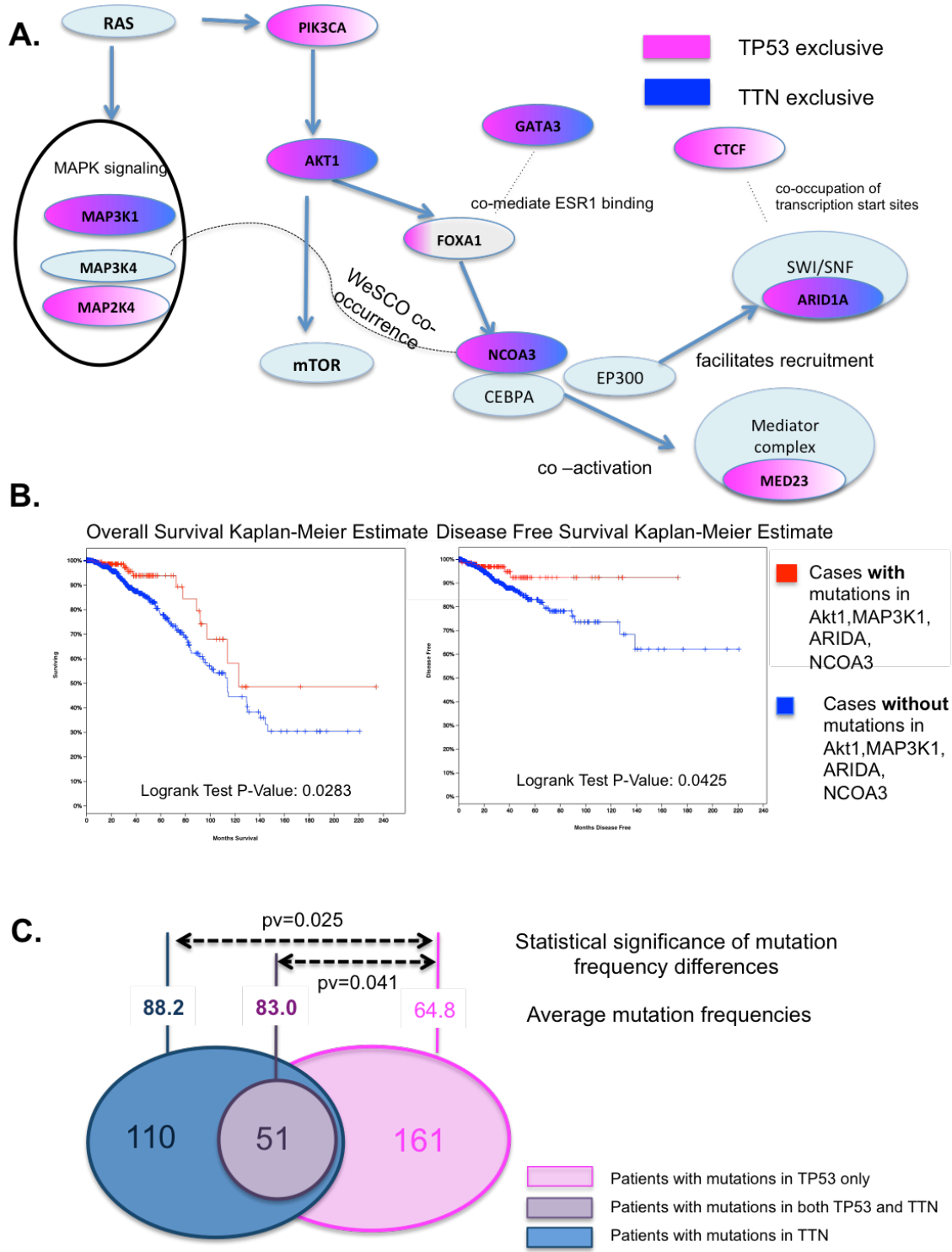

**Figure 4.** Mutual exclusivity of TTN and TP53 defines a group of related genes that are predictive of survival. A) Genes mutually exclusive with TP53 (purple) and TTN (blue) with FDR cutoff of 0.125 (except of FOXA1 for which p-value is 0.00365 and FDR = 0.3) with their context of functional interactions based on [31]. The interactions between FOXA1 and GATA3 are based on [32], and the relationships between EP300 and SWI/SNF and between CTCF and SWI/SNF are from [33, 34] and [35], respectively. B) Cancer driver genes AKT1, NCOA3, ARID1A, and MAP3K1 that mutually exclusive with both TTN and TP53 provide a better survival prognosis. C) Comparison of mutation frequencies for patients with TTN and TP53 mutations. The numbers in the Venn diagram represent the number of patients and the average mutation frequencies of the corresponding groups are shown above the diagrams.

## ACKNOWLEDGEMENTS

This work was supported in part by the Intramural Research Program of the National Institutes of Health, National Library of Medicine

## REFERENCES

1. Kim, Y.A., et al., *MEMCover: integrated analysis of mutual exclusivity and functional network reveals dysregulated pathways across multiple cancer types.* Bioinformatics, 2015. **31**(12): p. i284-92.
2. Ciriello, G., et al., *Mutual exclusivity analysis identifies oncogenic network modules.* Genome Res, 2012. **22**(2): p. 398-406.
3. Leiserson, M.D., et al., *Simultaneous identification of multiple driver pathways in cancer.* PLoS Comput Biol, 2013. **9**(5): p. e1003054.
4. Thomas, R.K., et al., *High-throughput oncogene mutation profiling in human cancer.* Nat Genet, 2007. **39**(3): p. 347-51.
5. Vandin, F., E. Upfal, and B.J. Raphael, *De novo discovery of mutated driver pathways in cancer.* Genome Res, 2012. **22**(2): p. 375-85.
6. Leiserson, M.D., et al., *CoMEt: a statistical approach to identify combinations of mutually exclusive alterations in cancer.* Genome Biol, 2015. **16**: p. 160.
7. Kim, Y.A., D.Y. Cho, and T.M. Przytycka, *Understanding Genotype-Phenotype effects in Cancer via Network Approaches.* PLoS Comput Biol, 2016(In Print).
8. Babur, O., et al., *Systematic identification of cancer driving signaling pathways based on mutual exclusivity of genomic alterations.* Genome Biol, 2015. **16**: p. 45.
9. Hofree, M., et al., *Network-based stratification of tumor mutations.* Nat Methods, 2013. **10**(11): p. 1108-15.
10. Kim, Y.A. and T.M. Przytycka, *Bridging the Gap between Genotype and Phenotype via Network Approaches.* Front Genet, 2012. **3**: p. 227.
11. Vogelstein, B. and K.W. Kinzler, *Cancer genes and the pathways they control.* Nat Med, 2004. **10**(8): p. 789-99.
12. Szczurek, E. and N. Beerenwinkel, *Modeling mutual exclusivity of cancer mutations.* PLoS Comput Biol, 2014. **10**(3): p. e1003503.
13. Kandoth, C., et al., *Mutational landscape and significance across 12 major cancer types.* Nature, 2013. **502**(7471): p. 333-9.
14. Wang, X., et al., *Widespread genetic epistasis among cancer genes.* Nat Commun, 2014. **5**: p. 4828.
15. Dwyer-Nield, L.D., et al., *Epistatic interactions govern chemically-induced lung tumor susceptibility and Kras mutation site in murine C57BL/6J-ChrA/J chromosome substitution strains.* Int J Cancer, 2010. **126**(1): p. 125-32.
16. Lawrence, M.S., et al., *Mutational heterogeneity in cancer and the search for new cancer-associated genes.* Nature, 2013. **499**(7457): p. 214-8.
17. Harris, R.S., *Molecular mechanism and clinical impact of APOBEC3B-catalyzed mutagenesis in breast cancer.* Breast Cancer Res, 2015. **17**: p. 8.
18. Burns, M.B., et al., *APOBEC3B is an enzymatic source of mutation in breast cancer.* Nature, 2013. **494**(7437): p. 366-70.
19. Alexandrov, L.B., et al., *Signatures of mutational processes in human cancer.* Nature, 2013. **500**(7463): p. 415-21.
20. Shannon, P., et al., *Cytoscape: a software environment for integrated models of biomolecular interaction networks.* Genome Res, 2003. **13**(11): p. 2498-504.
21. Su, Z.Z., et al., *PEG-3, a nontransforming cancer progression gene, is a positive regulator of cancer aggressiveness and angiogenesis.* Proc Natl Acad Sci U S A, 1999. **96**(26): p. 15115-20.
22. Banerji, S., et al., *Sequence analysis of mutations and translocations across breast cancer subtypes.* Nature, 2012. **486**(7403): p. 405-9.
23. Sun, X.X., et al., *The nucleolar ubiquitin-specific protease USP36 deubiquitinates and stabilizes c-Myc.* Proc Natl Acad Sci U S A, 2015. **112**(12): p. 3734-9.
24. Lawrence, M.S., et al., *Discovery and saturation analysis of cancer genes across 21 tumour types.* Nature, 2014. **505**(7484): p. 495-501.
25. Cancer Genome Atlas Research, N., et al., *Integrated genomic characterization of endometrial carcinoma.* Nature, 2013. **497**(7447): p. 67-73.
26. Cescon, D.W., B. Haibe-Kains, and T.W. Mak, *APOBEC3B expression in breast cancer reflects cellular proliferation, while a deletion polymorphism is associated with immune activation.* Proc Natl Acad Sci U S A, 2015. **112**(9): p. 2841-6.
27. Periyasamy, M., et al., *APOBEC3B-Mediated Cytidine Deamination Is Required for Estrogen Receptor Action in Breast Cancer.* Cell Rep, 2015. **13**(1): p. 108-21.
28. Wang, Y., et al., *The role of innate APOBEC3G and adaptive AID immune responses in HLA-HIV/SIV immunized SHIV infected macaques.* PLoS One, 2012. **7**(4): p. e34433.
29. Henderson, S., et al., *APOBEC-mediated cytosine deamination links PIK3CA helical domain mutations to human papillomavirus-driven tumor development.* Cell Rep, 2014. **7**(6): p. 1833-41.
30. Vieira, V.C. and M.A. Soares, *The role of cytidine deaminases on innate immune responses against human viral infections.* Biomed Res Int, 2013. **2013**: p. 683095.
31. Tamborero, D., et al., *Comprehensive identification of mutational cancer driver genes across 12 tumor types.* Sci Rep, 2013. **3**: p. 2650.
32. Theodorou, V., et al., *GATA3 acts upstream of FOXA1 in mediating ESR1 binding by shaping enhancer accessibility.* Genome Res, 2013. **23**(1): p. 12-22.
33. Dallas, P.B., et al., *p300/CREB binding protein-related protein p270 is a component of mammalian SWI/SNF complexes.* Mol Cell Biol, 1998. **18**(6): p. 3596-603.
34. Ogiwara, H., et al., *Histone acetylation by CBP and p300 at double-strand break sites facilitates SWI/SNF chromatin remodeling and the recruitment of non-homologous end joining factors.* Oncogene, 2011. **30**(18): p. 2135-46.
35. Euskirchen, G.M., et al., *Diverse roles and interactions of the SWI/SNF chromatin remodeling complex revealed using global approaches.* PLoS Genet, 2011. **7**(3): p. e1002008.

# Supplementary Materials
# WeSME: Uncovering Mutual Exclusivity of Cancer Drivers and Beyond

Yoo-Ah Kim[1], Sanna Madan[2], and Teresa M. Przytycka[1] *

[1] National Center for Biotechnology Information, NLM, NIH, Bethesda, MD 20894

[2] Poolesville High School, Poolesville, MD

*To whom correspondence should be addressed; przytyck@ncbi.nlm.nih.gov

## A   COMPARISON OF WESME AND PERMUTATION METHOD

The main difference between WeSME and the traditional permutation method lies in how random mutation profiles are generated. Once random profiles are generated, the computation of empirical p-values can be performed in a similar way for both methods. To compare their computational cost, we measured the total time and storage required to generate random profiles using the two methods for the somatic mutation profile of 665 TCGA BRCA patients (Table S1). We used all genes with at least one mutation without pre-filtering. In the permutation method, a mutation matrix is shuffled while the mutation frequencies of cancer patients and the mutation rates of genes are preserved*. For the purpose of comparison, we show the cost when WeSME samplings were run for all genes (the second column in Table S1) but note that we only need to generate random profiles for each different $k$ (the number of mutated patients) instead of generating them for all genes (see Section 2.2). The last column in Table S1 shows the results for compact sampling.

|  |  | permutation | WeSME for all genes | WeSME |
|---|---|---|---|---|
| 1 random profile | time (sec) | 100.67 | 2.31 | 0.018 |
|  | file size (KB) | 226 | 226 | 1.2 |
| $10^4$ random profiles[†] | total time (hour) | 279.6 | 6.4 | 0.05 |
|  | total file size (GB) | 2.26 | 2.26 | 0.012 |

**Table S1:** Comparison of the total computation time and storage required for the permutation method and WeSME. For WeSME, the data have been measured in two scenarios – i) sampling is performed for each gene separately and ii) sampling only for each different number of mutated patients.

---

* To ensure the randomness of the data, we performed swapping of mutations $100*|E|$ times where $|E|$ is the total number of mutations [1]

[†] $10^4$ random profiles correspond to the p-value precision depth of $10^{-4}$. While the permutation method requires generating $10^p$ random instances for an entire mutation profile to obtain the precision of $10^{-p}$ in p-values, WeSME can decide the precision depth of p-values independently for different gene pairs. We show the time and storage requirement to generate the same $10^4$ random instances using WeSME only for the purpose of comparison.

Besides the reduced computation time and file sizes, an important advantage of WeSME is that the samplings for each gene can be generated separately *as needed*. For instance, in case we are interested in finding out the relationships among a certain subset of genes, random weighted samplings can be generated only for the genes of interest quickly whereas the permutation method requires performing permutations for the entire genome. This property also allows WeSME dynamically to adjust the precision of empirical p-values by generating more random profiles for some gene pairs if necessary. In the permutation method, $10^P$ permuted matrices of the whole genome should be generated to obtain the precision of $10^{-P}$ even though such precision may not be necessary for most of non-significant gene pairs.

## B  SLIGHT TENDENCY OF WESME TO OVERESTIMATE P-VALUES FOR HIGHLY MUTATED GENES

We noticed in Figure 2A that WeSME slightly overestimated p-values compared to the permutation method especially for highly mutated genes. We conjecture that this is due to the fact that weighted samplings were performed for each gene (or for each different $k$ more precisely) independently. In the permutation test, permutations are performed genome-wide and there exists dependency between genes to keep the mutation frequencies preserved. WeSME uses the mutation frequencies of patients as weights for sampling to ensure that each patient is sampled with the probability in proportion to its mutation frequency, but the mutation frequency in null distribution is not guaranteed.

To find out how the mutation frequencies differ in the null distributions of two methods, we plotted the patient mutation frequencies in the generated null distribution for PIK3CA (y-axis in Figure S1 (A)) and for TTN (Figure S2 (A)) along with their observed mutation frequency from the original mutational profile (x-axis). PIK3CA is the most highly mutated genes in the dataset (237 out of 665 BRCA patients) and TTN is mutated in 110 patients. To obtain the null frequency of each patient, we generated 1,000 random mutation profiles using both methods and then computed, for each patient, the fraction of instances in which the patient has a mutation in PIK3CA (or TTN). The observed mutation frequency of a patient $s$ (x-axis) is obtained by computing $mf(s) * m(g)$ with an assumption of independent sampling (i.e., sampling $m(g)$ times with probability $mf(s)$ with replacement). In methods, the null frequencies and observed mutation frequencies are not linear especially for the highly mutated patients because the random samplings or permutations do not allow replacements (therefore the null frequency cannot be greater than 1). As we can see in the figure, the patients with higher mutation frequencies tend to have higher null frequencies in WeSME than in permutation method. We believe that this is because the permutation test ensures the mutation frequency of each patient remains the same when generating random profiles and therefore, the extra dependency creates less frequency

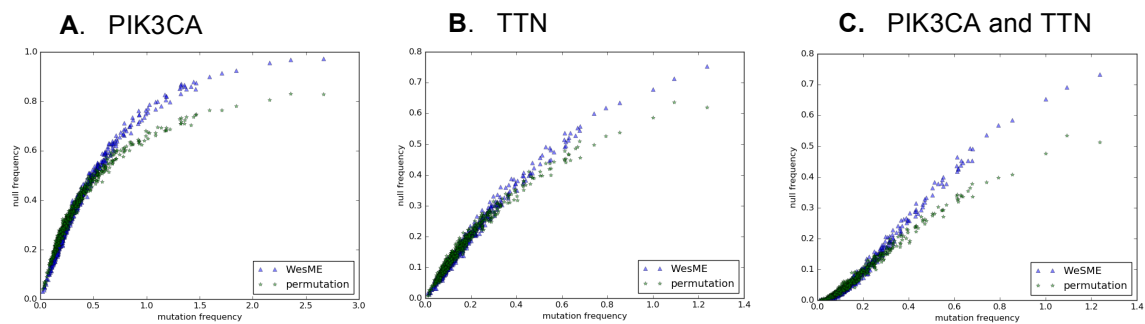

**Figure S1.** Comparison of mutation frequencies of each patient in the null distributions generated by WeSME and permutation (A) mutations in PIK3CA (B) mutations in TTN (C) mutations in both PIK3CA and TTN

variations among patients.

The ME p-values between PIK3CA and TTN are 0.247 and 0.61 using WeSME and the permutation test, respectively. To investigate the underlying cause of the difference, we plotted, for each patient, the fraction of instances (out of 1,000 random mutation profiles) in which both PIK3CA and TTN have mutations in the patient (Figure S1(C)). Each dot represents a patient with its null frequency of co-occurrence in y-axis. While WeSME and permutation methods have similar co-occurring frequencies in the patients with low mutation frequencies, the difference is more noticeable in the highly mutated patients. This implies that the two genes are more likely to co-occur in WeSME null distribution than in the null distribution generated by permutations, which in turn translates to more significant p-values in WeSME compared to the permutation method. We do not observed similar patterns in less mutated genes and the p-values tend to be similar in both methods for those genes.

## C  WESCO VALIDATION

As with WeSME, we compared the p-values of co-occurrence with WeSCO, permuation and hypergeometric test and plotted the results in Figure S2. Due to the time required to compute p-values using permutation and hypergeometric test, we only considered 1% randomly selected pairs with WeSCO p-value < 0.25. As shown in the figure, WeSCO p-values correlates very well with the p-values computed with the permutation test whereas the p-values obtained using hypergeometric test tends to overestimate co-occurrence.

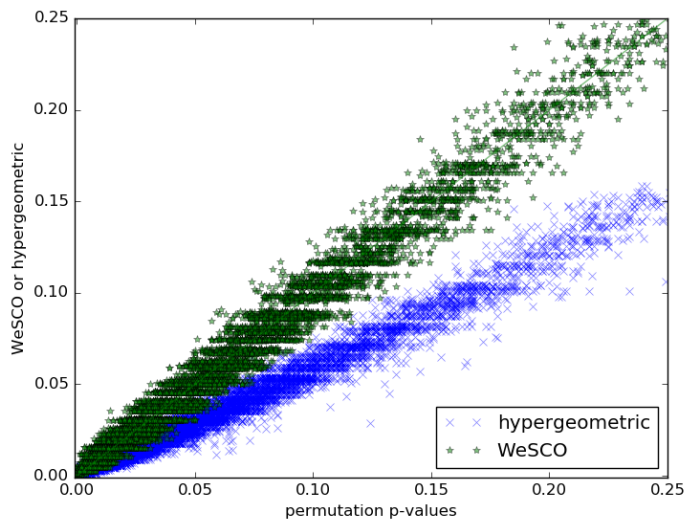

**Figure S2.** Comparison of the p-values computed based on permutation, hyergometric and WeSCO tests. Only the data for 1% of random chosen pairs with their p-values ≤ 0.25 are shown.

## D  WESME PSEUDOCODE

---

**Algorithm 1. Compute the weights of patients**

**Input** binary mutation matrix M

**Output** mutation frequency vector $(mf(s_1), mf(s_2) \ldots mf(s_n))$

1. for each patient $s_i$:
   $mc(s_i) \leftarrow$ the number of genes mutated in $s_i$
2. compute the mutation frequency vector $(mf(s_1), mf(s_2) \ldots mf(s_n))$ by normalizing $(mc(s_1), mc(s_2) \ldots mc(s_n))$ with the total number of mutations $\Sigma_i\, mc(s_i)$
3. return $(mf(s_1), mf(s_2) \ldots mf(s_n))$

---

**Algorithm 2.  Generate weighted sampling**

**Input**  mutation frequency vector $(mf(s_1), mf(s_2) \ldots mf(s_n))$
      $k$: number of patients to be sampled each time
      $W$: number of times to be performed sampling

**Output** $W$ sets of $k$ random patients $WS(k) = (R^1, R^2, R^3 \ldots R^W)$, $|R^j| = k$

1. for $j = 1 \ldots W$:
   $R^j \leftarrow$ randomly select $k$ patients without replacement so that the probability to choose $s_i$ is $mf(s_i)$
2. return $(R^1, R^2, R^3 \ldots R^W)$

---

**Algorithm 3.  Compute ME p-value**

**Input**  $g_1, g_2$: gene pair
      $W_o$: initial number of sampling pairs
      $W_m$: maximum number of sampling pairs
      $p_{min}$: minimum number of sampling pairs with bigger *ExCover*

**Output**  $pv$: $p$-value

1. $S(g_i) \leftarrow$ set of patients with $g_i$ mutated ($i = 1, 2$)
2. $OriginalCover \leftarrow ExCover(S(g_1), S(g_2))$
3. $k_i \leftarrow m(g_i)$, ($i = 1, 2$)
4. $W \leftarrow W_o$
5. while $W \leq W_m$:
   $P \leftarrow$ Randomly selected $W$ pairs of indices $= ((l_1^1, l_2^1) \ldots (l_1^W, l_2^W))$
       s.t. $l_1^j \neq l_2^j$ for all $j$ and $(l_1^j, l_2^j) \neq (l_1^k, l_2^k)$ for any $j \neq k$
   $pv = 0$
   For each pair $(l_1^j, l_2^j)$ in $P$: // compute p-value
       $R_i \leftarrow l_i^j$-th set in $WS(k_i)$, ($i = 1, 2$)
       If $ExCover(R_1, R_2)) \geq OriginalCover$:
           $pv \leftarrow pv + 1/W$
   if $pv \leq p_{min}/W$:  // increase precision
       $W \leftarrow W*10$
6. return $pv$

**Algorithm 4. Compute FDR**

**Input** $M$: binary mutation matrix
  $PVS$: WeSME p-values for all gene pairs
  $mth$: mutation rate threshold for bins

**Output** $FDRS$: FDRs of the gene pairs

1. for $t$ in $[1..100]$: // *compute WeSME p-values with permuted instances*
    $M_t \leftarrow$ permute $M$
    For each gene pair $(g_i, g_j)$:
      $PV_t(g_i, g_j) \leftarrow$ compute WeSME p-value based on $M_t$
   // *use three bins for gene pairs by dividing genes into two groups*
   // *can be extended to use more bins if necessary*
2. $R \leftarrow$ genes with mutation rate $\leq mth$
3. $H \leftarrow$ genes with mutation rate $> mth$
4. divide gene pairs into three bins $[(R, R), (R, H), (H, H)]$
5. for each bin $B$ in $[(R, R), (R, H), (H, H)]$:
    $PVS_B \leftarrow$ PVS for gene pairs in $B$
    $PVS_B^{null} \leftarrow$ empty list
    For $t$ in $[1..100]$:
      $PVS_B^{null} \leftarrow PVS_B^{null} + PV_t$ for gene pairs in $B$
    For each gene pair $(g_i, g_j)$ in $B$:
      $V =$ rank of $PVS(g_i, g_j)$ in $PVS_B^{null}/|PVS_B^{null}|$
      $R =$ rank of $PVS(g_i, g_j)$ in $PVS_B/|PVS_B|$
      $FDRS(g_i, g_j) = V/R$
6. return FDRS

## E  TCGA ACUTE MYELOID LEUKEMIA DATASET ANALYSIS

We applied WeSME in 194 TCGA Acute Myeloid Leukemia patients [2] and the gene pairs with FDR < 0.15 are shown in Figure S3. We used two bins with 5% mutation rate cutoff.

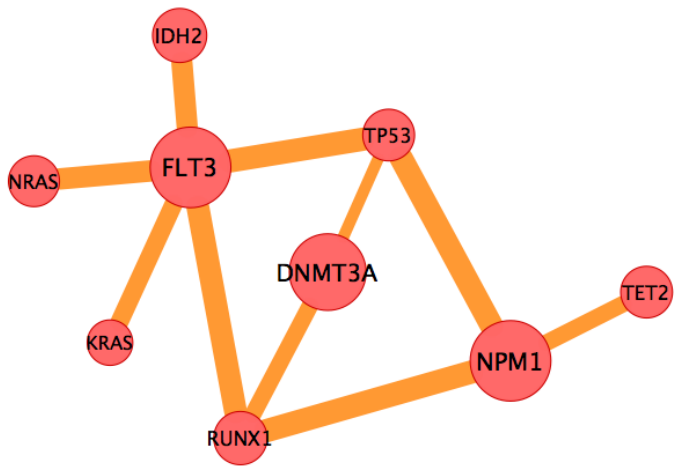

**Figure S3.** ME network for TCGA LAML (FDR < 0.15). The node size of each gene represents the mutation rate of the gene for LAML. The widths of edges correspond to their WeSME p-values (thicker edges for more significant pairs) whereas darker colors of the edges correspond to lower FDRs. The networks are created using Cytoscape [3]

1. Milo, R., et al., *On the uniform generation of random graphs with prescribed degree sequences*. 2003, arXiv:cond-mat/0312028.
2. Cancer Genome Atlas Research, N., *Genomic and epigenomic landscapes of adult de novo acute myeloid leukemia*. N Engl J Med, 2013. **368**(22): p. 2059-74.
3. Shannon, P., et al., *Cytoscape: a software environment for integrated models of biomolecular interaction networks*. Genome Res, 2003. **13**(11): p. 2498-504.